\documentclass[fleqn,usenatbib]{mnras}
\usepackage{newtxtext,newtxmath}
\usepackage[T1]{fontenc}
\DeclareRobustCommand{\VAN}[3]{#2}
\let\VANthebibliography\thebibliography
\def\thebibliography{\DeclareRobustCommand{\VAN}[3]{##3}\VANthebibliography}
\usepackage{graphicx}	% Including figure files
\usepackage{amsmath}	% Advanced maths commands
\usepackage{wasysym}
\usepackage{multicol}        % Multi-column entries in tables
\usepackage{bm}		% Bold maths symbols
\usepackage{pdflscape}	% Landscape pages
\usepackage{gensymb}
\usepackage{hyperref}

\title{Doppler Shifted Transient Sodium Detection by KECK/HIRES}

\author[Athira Unni et al.]{
Athira Unni,$^{1,2}$\thanks{E-mail: athira.exo@gmail.com}
Apurva V. Oza,$^{3,4}$ H. Jens Hoeijmakers,$^{5}$ Julia V. Seidel,$^{6}$ Thirupathi Sivarani,$^{2}$
\newauthor
Carl A. Schmidt,$^{7}$ Aurora Y. Kesseli,$^{8}$ Katherine de Kleer,$^{3}$ 
Ashley D. Baker,$^{9}$ Andrea Gebek, 
\newauthor
$^{10}$ Moritz Meyer zu Westram,$^{11}$ Chloe Fisher,$^{12}$   Steph Sallum, $^{1}$ Manjunath Bestha, $^{2}$ and Aaron Bello-Arufe $^{4}$ \\
$^{1}$ Department of Astronomy and Astrophysics, University of California, Santa Cruz, USA\\
$^{2}$Indian Institute of Astrophysics, Bangalore, India\\
$^{3}$ Division of Geological and Planetary Sciences, California Institute of Technology, Pasadena, USA\\
$^{4}$Jet Propulsion Laboratory, California Institute of Technology, Pasadena, USA\\
$^{5}$ Department of Astronomy and Theoretical Physics, Lund Observatory, Lund, Sweden\\
$^{6}$ European Southern Observatory, Santiago de Chile, Chile\\
$^{7}$ Center for Space Physics, Boston University, Boston, USA\\
$^{8}$ IPAC, California Institute of Technology, Pasadena, USA\\
$^{9}$Department of Astronomy, California Institute of Technology, Pasadena, USA\\
$^{10}$Sterrenkundig Observatorium, Universiteit Gent, Ghent, Belgium\\
$^{11}$ Astronomical Institute, University of Bern,  Bern, Switzerland\\
$^{12}$ Department of Physics, University of Oxford, Oxford, UK \\
}

\date{Accepted XXX. Received YYY; in original form ZZZ}
\pubyear{2023}

\begin{document}
\label{firstpage}
\pagerange{\pageref{firstpage}--\pageref{lastpage}}
\maketitle
\begin{abstract}
% 200 words
We carried out the first high-resolution transit observations of the exoplanet WASP-49 Ab with Keck/HIRES. Upon custom wavelength calibration we achieve a Doppler RV precision of $<$ 60 {\mbox{${\rm m\,s}^{-1}$}}. This is an improvement in RV stability of roughly 240 {\mbox{${\rm m\,s}^{-1}$}} with respect to the instrument standard.
We report an average sodium flux residual of $\Delta \mathcal{F}_{NaD}/ \mathcal{F}_{\star} (\lambda)  \sim$  3.2 $\pm$ 0.4 $\%$ (8.0 $\sigma$) comparable to previous studies. Interestingly, an average Doppler shift of -6.2 $\pm$ 0.5 {\mbox{${\rm km\,s}^{-1}$}} (12.4 $\sigma$) is identified offset from the exoplanet rest frame. The velocity residuals \textit{in time} trace a blueshift (v$_{\Gamma, ingress} \sim$ -10.3 $\pm$ 1.9 {\mbox{${\rm km\,s}^{-1}$}}) to redshift (v$_{\Gamma, egress} \sim$ +4.1 $\pm$ 1.5 {\mbox{${\rm km\,s}^{-1}$}}) suggesting the origin of the observed sodium is unlikely from the atmosphere of the planet. The average Na light curves indicate a depth of $\Delta \mathcal{F}_{NaD} /\mathcal{F}_{\star} (t) \sim$ 0.47 $\pm$ 0.04 \% (11.7 $\sigma$) enduring $\lesssim$ 90 minutes with a half-max duration of $\sim$ 40.1 minutes.  Frequent high-resolution spectroscopic observations will be able to characterize the periodicity of the observed Doppler shifts. Considering the origin of the transient sodium gas  is of unknown geometry, a co-orbiting natural satellite may be a likely source.
\end{abstract}

\begin{keywords}
techniques: spectroscopic  -- Exoplanets -- planets and satellites: atmospheres  -- planets and satellites: gaseous planets  -- planets and satellites: composition
\end{keywords}

\section{Introduction} \label{intro_wasp49}
High-Resolution Transmission Spectroscopy (HRTS) is a powerful tool for exoplanet atmospheric characterization. High spectral resolution allows us to resolve and detect individual atomic and molecular species without the help of atmospheric retrieval models \citep{Snellen_2008}. It is also possible to detect signatures of global circulation and contamination from stellar chromospheric emission. The first high-resolution exoplanet transit spectra were successfully obtained for HD189733b \citep{Redfield_2008} using the High-Resolution Spectrograph at the Hobby-Eberly Telescope. Apart from this, VLT/ESPRESSO \citep{pepe_2013}, HARPS-N/GIANO \citep{cosentino_2012,claudi_2017} at the Telescopio Nazionale Galileo (TNG), CARMENES \citep{quirrenbach_2010} at Calar Alto Observatory, and SPIRou \citep{mclean_2012} at the Canada France Hawaii Telescope (CFHT) showed their capability of HRTS observations. Sodium is considered to be a commonly studied element in the optical part of the transmission spectra of hot Jupiters and hot Saturns \citep{Redfield_2008, Snellen_2008,wood_2010, Zhou_2012,Burton_2015, Wyttenbach_2015, Wyttenbach2017, Jensen_2018, Seidel2020b, Seidel2020c,Allart_2020,Chen_2020,Ishizuka_2021,Kawauchi2021}. Due to the large resonance scattering cross-section, sodium allows for detections at high altitudes where the annulus of gas is expected to be rarefied and tenuous, in the planet's alkali exosphere \citep{Gebek_2020}. 

Here we carefully study the Na D line region of WASP-49 Ab during the transit using the high-resolution transit observation from Keck/HIRES. WASP-49 Ab orbits a G6V star (Vmag=11.35, distance=194.5 pc) with a transit duration of 2.14 hours and a mass of 0.37$M_{J}$ \citep{Lendl2012}. The first, high-resolution transmission spectra of WASP-49 Ab were obtained using HARPS/3.6-m spectrograph by \citet{Wyttenbach2017},  revealing anomalous sodium of $dF_{Na}/F_{\star} (\lambda) \approx 1.99 \% $ and $1.83 \% $ for D$_2$ and D$_1$ lines, respectively. The observed neutral sodium was reported to be extended up to $\sim$ 1.5 times the planetary radii to reproduce the observed sodium transit depth. This is roughly three times the Na I altitude probed at any known exoplanetary system to date \citep{Langeveld_2022, Sicilia_2025}. 

Suppose we assume the origin of this sodium is endogenic, or from the atmosphere of the planet itself, then a super-solar sodium abundance or a super-heated atmospheric layer at WASP-49b may be plausible in reproducing the average altitude probed of neutral sodium. However, when atmospheric escape due to incoming irradiation is studied, the upper atmosphere cools and the line depth observed by HARPS/3.6-m could not be reproduced \citep{cubillos_2017}. Meanwhile, sodium in non-local thermodynamic equilibrium (N-LTE) was also suggested as a source \citep{fisher_heng2019}. In the case of an exogenic sodium, the possible sources can be an older cometary impact or a natural satellite orbiting around the planet WASP-49 A b \citep{oza_2019}.

Here we present the first Keck/HIRES observation of WASP-49b to understand the sodium anomaly. The paper is outlined as follows. In Sect \ref{section:2}, we
summarize the observations and the data reduction. A detailed wavelength re-calibration and analysis are presented in Sect \ref{section:wave}. Followed by the discussion and the results are presented in Sect. \ref{section:3}. Finally, Sect \ref{section:4} presents the conclusion.

%%%%%%%%%%%%%%%%%%%%%%%%%%%%%%%%%%%%%%%%%%%%%%%%%%%%%%%%%%%%
\section{Observation and data reduction} \label{section:2}
%%%%%%%%%%%%%%%%%%%%%%%%%%%%%%%%%%%%%%%%%%%%%%%%%%%%%%%%%%%%
HIgh Resolution Echelle Spectrograph (HIRES) is designed to be a general-purpose spectrograph installed on the Nasmyth platform of the Keck-I telescope. A wavelength range of 3000 - 10000 Å for the spectrograph is achieved using two different cross-dispersers and collimators. Three separate CCDs cover the entire echelle order of the spectra with an average spectral resolution of $\mathcal{R}$ = $67,000$ \citep{vogt_hires}.

One complete transit of WASP-49 Ab was observed with HIRES on 2019-11-11 (Program ID: C284).  The observation started from UT 10:28:41 (hh:mm:ss) to 16:15:05 with an exoplanet mid-transit time at 13.31 UT (Julian Date 2458799.06791922). A total of  42  exposures of 420 s each covered ingress, egress, complete transit, and out-of-transit phases. The sky was clear throughout the night. The typical SNR of the observation was 55 (at mid-transit). The airmass and the seeing varied from 1.30 to 1.76 and 0.6" to 1.2", respectively from the beginning to the end of the observation. We used the MAKEE pipeline reduced data from the Keck Observatory Archive (KOA) for the analysis. The continuum normalization of each order of all the spectra was performed using IRAF  with a cubic spline fitting.   

\section{Custom Wavelength Calibration} \label{section:wave} % 
The first successful attempt at using Keck/HIRES for the exoplanet atmospheric sodium detection was by \citet{Langland-Shula_2009} for the exoplanet HD 209458 b. As they were focused on detecting the broad-band, low-resolution sodium feature from the atmosphere of HD 209458 b rather than constructing high-resolution transit spectra, a linear shift using a cross-correlation in the wavelength solution was sufficient. HIRES is not housed inside a thermal or pressure-controlled environment; hence it is expected that there could be instrument drift that will cause a shift in the wavelength dispersion solution during the observations. Though slit spectrographs provide high throughput, the line spread function (LSF) can vary significantly during the observations. Due to this, the wavelength position on the detector shifts, and the wavelength dispersion solution could also change. The stability of the wavelength dispersion solution is crucial to perform HRTS observations. 

\begin{figure}
\begin{center}
\includegraphics[width=1.0\columnwidth]{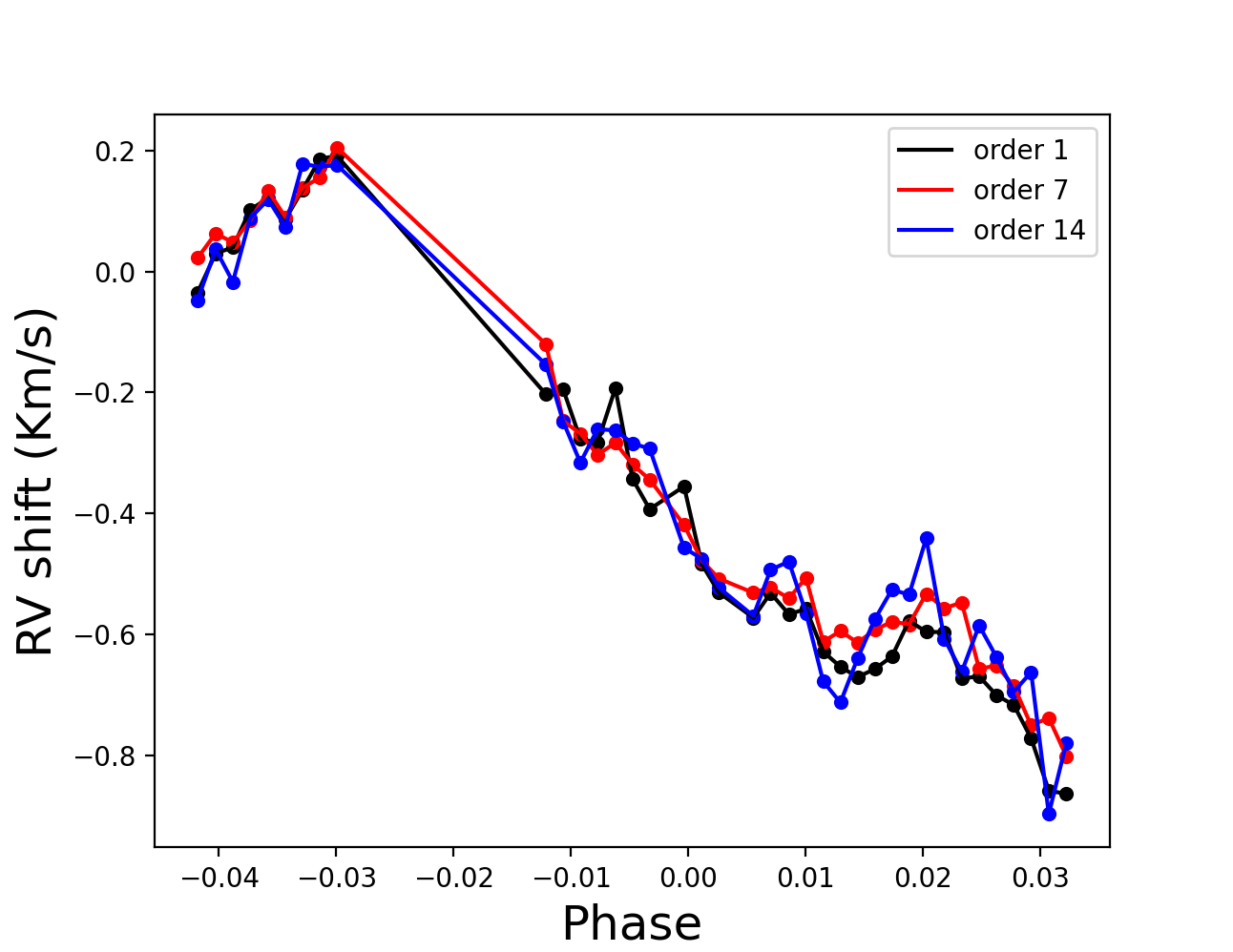}
\caption{The velocity shift in each exposure with respect to the reference exposure as a function of phase is shown for three echelle orders (HIRES CCD2). The first order is in black, the last order is in blue, and the order containing the sodium doublet around 5890 {\AA} is in red. There is a shift of 0.8 {\mbox{${\rm Km\,s}^{-1}$}} during the complete observation period.} \label{rv_correction1}
\end{center}
\end{figure}

To achieve the maximum wavelength precision, we used the first out-of-transit spectra as a reference and shifted all the exposures with respect to the reference exposure.  Figure \ref{rv_correction1} shows the velocity shift across the exposures over the complete observation period, which is about 0.8 {\mbox{${\rm km\,s}^{-1}$}}. We corrected this shift for individual exposures, which is significant for accurate telluric subtraction and detecting the residual transit velocity signal from the planet. 

\subsection{Distortion correction}
\begin{figure}
\centering
\includegraphics[width=\linewidth]{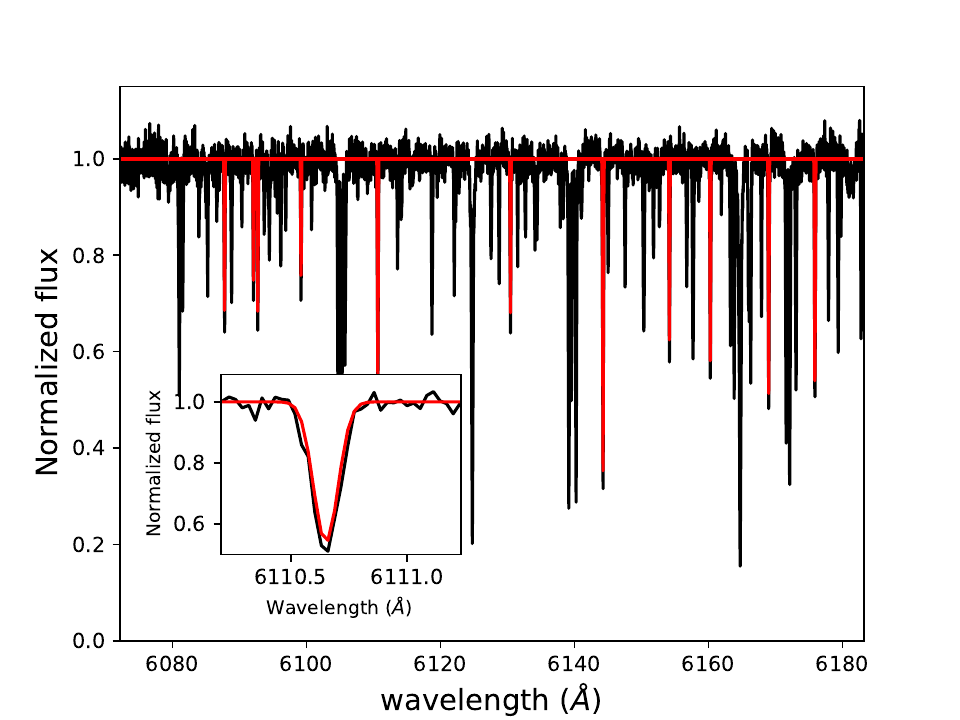}
\caption{ Observed spectra (in black) and selected clean stellar lines (red) for calculating the residual distortion. A Gaussian fit to the lines is shown in red. A zoomed version of a Gaussian fit is shown in the subplot.}\label{distortion_correction1}
\end{figure}

\begin{figure}
\centering
\includegraphics[width=\columnwidth]{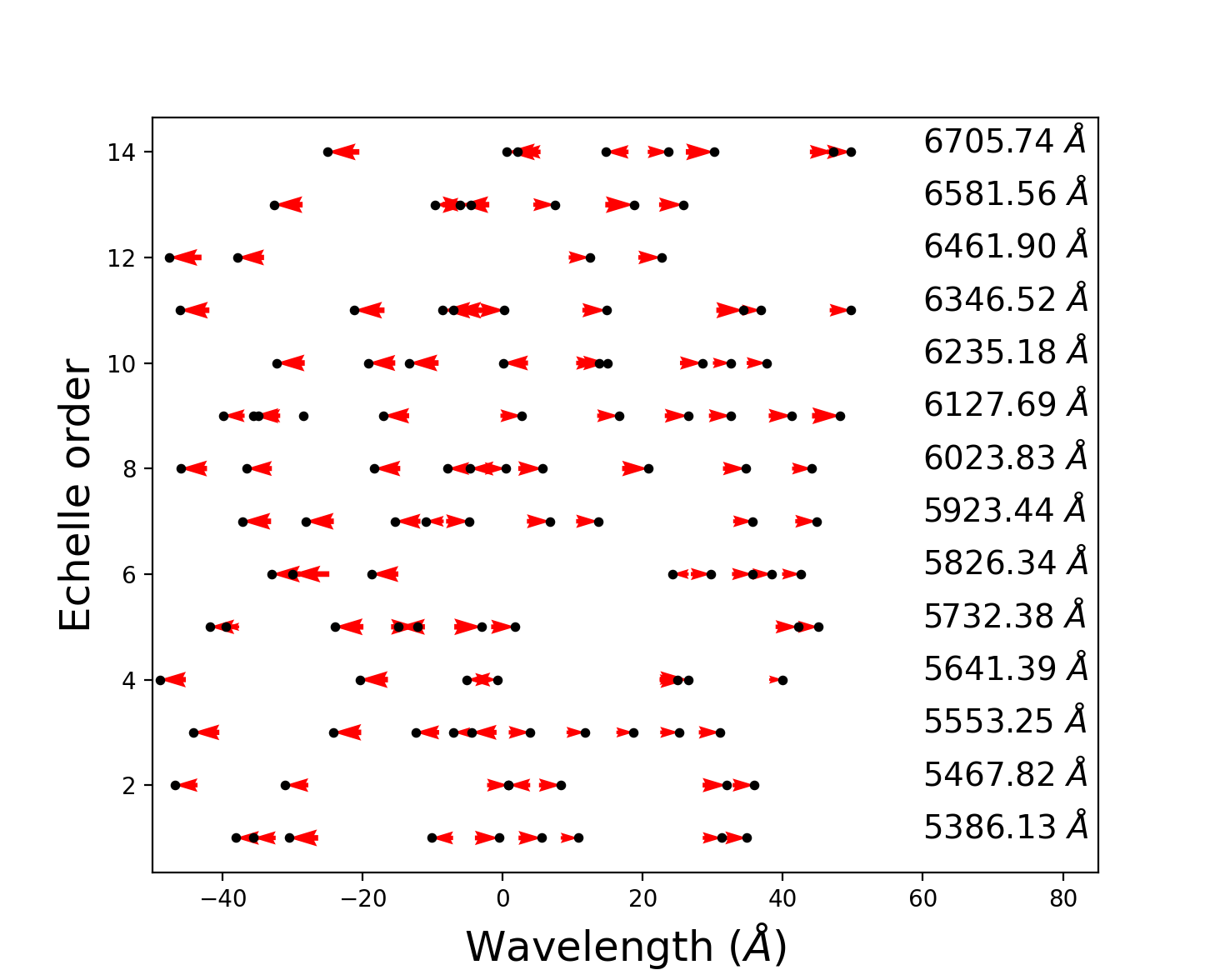}
\caption{A residual RV shift vector map across different echelle orders is shown here. The possible residual shift in each exposure is calculated by measuring the centroid of a set of lines in each exposure (Figure \ref{distortion_correction1}). The calculated shift in each line (0.0001 times length of the red arrow) with respect to the centroid of the corresponding line in the first exposure (black dot). The center wavelength of each order is considered to be zero. The y-axis represents the order number in KECK/HIRES CCD2. Also, the center wavelength of each order is mentioned on the right side of the plot. } \label{vector}
\end{figure}

The residual wavelength drift after correcting the linear velocity shift (Figure \ref{rv_correction1}) could arise due to optical distortion caused by aberration and varying anamorphism at the detector. And it is found to be $\pm$ 300 {\mbox{${\rm m\,s}^{-1}$}}. Ideally, re-deriving the wavelength dispersion solution will remove the error due to residual wavelength distortion. However, the ThAr calibration exposures were not taken during transit to maximize the transit coverage. So we used the stellar lines themselves to correct for the residual wavelength distortion errors separately for individual orders. For this purpose, we selected all the possible clean stellar lines from each order and found the wavelength centroid of these absorption lines using a Gaussian fit ( Figure \ref{distortion_correction1}). The shift in the centroid wavelength of each selected line, relative to the corresponding absorption line in the first exposure is estimated. These residual shift is not constant across the order and the magnitude of the residual distortion in wavelength was found to scale with overall instrument drift and minimum at the center of each order and increase as we go down to both edges of each order as shown in Figure \ref{vector}. A cubic spline was then fitted to these measured wavelength shifts in each order, relative to the reference measurements from the first exposure. Best-fit cubic spline coefficients in each order are used for wavelength re-calibration. Finally, we achieve a velocity precision of $\leq$ 60 {\mbox{${\rm m\,s}^{-1}$}} (Figure \ref{distortion} in Appendix)  for the sodium doublet order. This is the minimum shift that can be achieved using one-dimensional extracted spectra from Keck/HIRES. However, the first exposure itself can have a significant distortion. Therefore, we cross-correlate each exposure with respect to the synthetic solar spectra of the same spectral resolution, bringing all exposures to Stellar Rest Frame (C.f. Figure \ref{spectra_overplot}).

%%%%%%%%%%%%%%%%%% vector field plot %%%%%%%%%%%%%%%%%
\section{Results} \label{section:3}
%\subsection{Final transmission spectra}  
We followed the steps outlined in \cite{Seidel2019} to obtain the final transmission spectrum. Telluric correction was performed on the normalized spectra using {\it molecfit} \citep{Smette2015, Kausch2015}, version 1.5.1. {\it Molecfit} is a well-established tool, successfully used over the years by, e.g., \citet{Allart2017, Hoeijmakers2020, Seidel2020b, Seidel2020c, Kawauchi2021, Stangret2021}. All telluric corrected spectra in the Stellar Rest Frame (SRF) are separated into exposures taken during transit and out-of-transit, based on the transit ingress ($\phi_{exoplanet} = -0.016$) and egress phases ($\phi_{exoplanet} = 0.016$). Finally, we obtained 17 exposures taken during the transit and 20 spectra taken outside the transit event, after removing a couple of exposures with lower SNR. 

\begin{figure}%[htp]
\centering
\includegraphics[width=1.0\columnwidth]{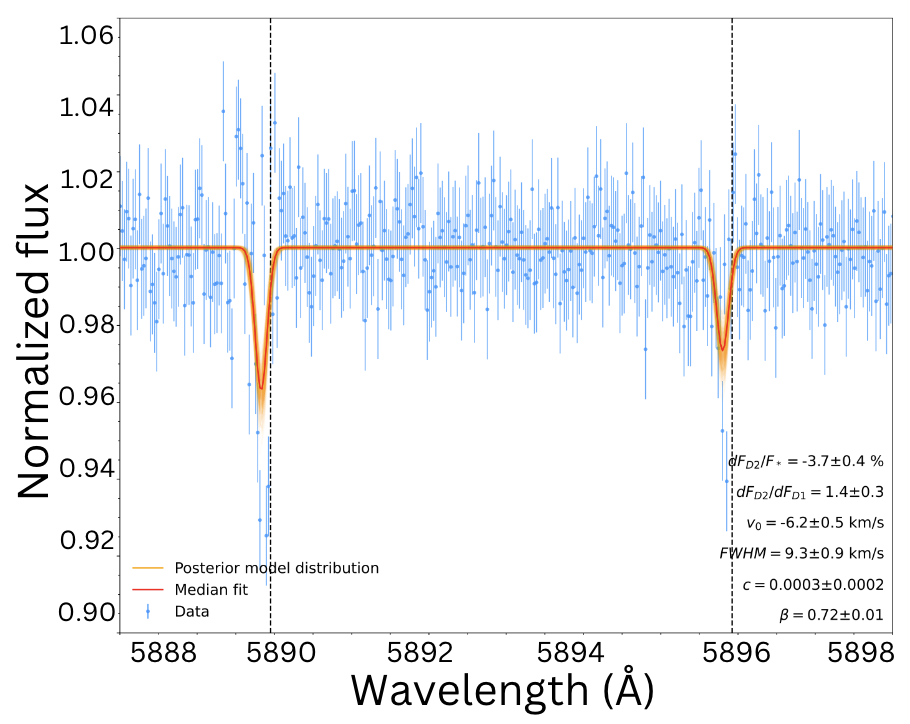}
\caption{Final transmission spectrum from KECK/HIRES best-fit Gaussian model along with the raw data. The average amplitude of the best-fit Gaussian is $dF_{Na}/F_{*}=$ 3.2$\pm$0.4 \% (8.0-$\sigma$), FWHM =  $9.3 \pm 0.9$   ${\mbox{${\rm km\,s}^{-1}$}}$ and continuum c. Uncertainties are scaled with $\beta$ as a free-fitting parameter. The residual sodium signal is blue shifted at $v_{0} =$ -6.2 $\pm$0.4 {\mbox{${\rm km\,s}^{-1}$}}.}
\label{masked_residual}
\end{figure}

\begin{figure}
\centering
\includegraphics[width=0.8\columnwidth]{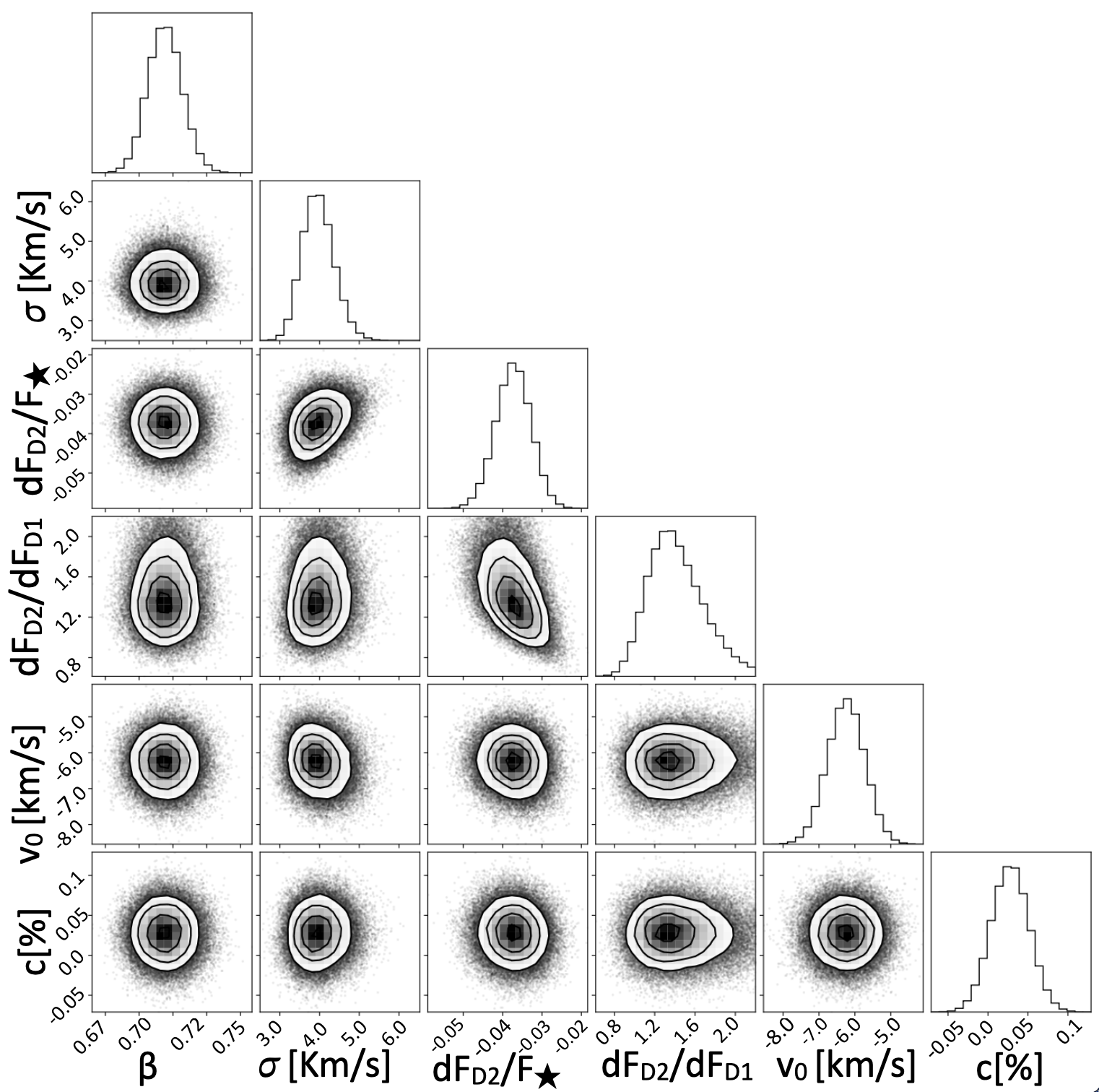}
\caption{Posterior distributions of the model parameters describing
the Na doublet as a pair of Gaussian where the average D$_2$ and D$_1$ line depth is reported $dF_{Na}/F_{*}$, centroid velocity shift $v_{0}$, width $ \sigma$, sodium $D_{2}$ to $D_{1}$ line ratio $dF_{D2}/dF_{D1}$, and continuum c. The uncertainties are scaled with $\beta$ which is a free fitting parameter.}
\label{GaussianEpoch5}
\end{figure}

\begin{figure}
\centering
%\hspace*{-4cm}
  \centering  \includegraphics[width=0.5\textwidth]{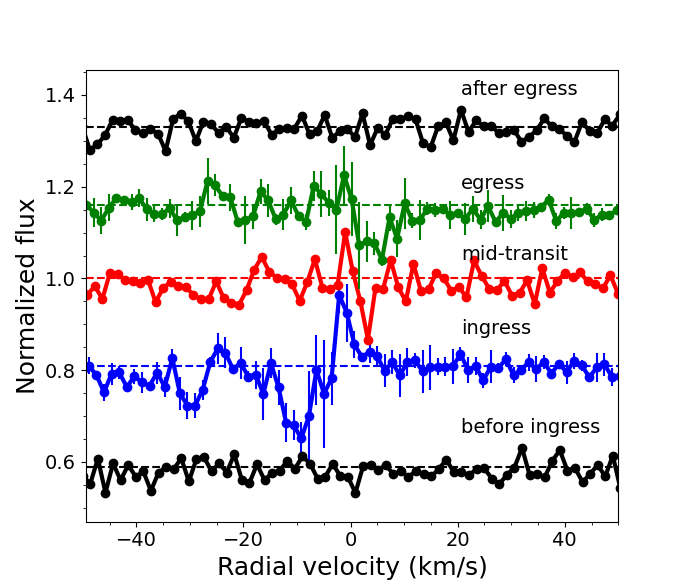} 
  \caption{Doppler shifts versus exoplanet phase for the NaD$_2$ line. Blue indicates ingress, red indicates mid-transit, and green indicates the egress part of the transit. Evidently, the largest residual peak is indeed blue shifted as described in the text. At egress, there is a significant redshift. A constant y-offset is applied for visibility purposes. We note that the Rossiter-Mclaughlin effect cannot generate signals larger than 100 ppm \citep{Wyttenbach2017}.}
  %\textbf{Julia, Athira, would it be useful to indicate the planetary limb Doppler shift ($\sim \pm$ 3 km/s?) and stellar atmosphere shift ( ($\sim \pm$ 5 km/s?) }}   
  \label{DopplerCC}
\end{figure}

%%%%%%%%%%%%%%%%%%%%%%%%%%%%%%%%%%%%%%%%%%%%%%
The residual spectra are obtained by dividing each in-transit spectra by the master out-of-transit spectra. These residuals are then shifted back to the Planet Rest Frame (PRF) by applying the estimated orbital velocity of the planet at each exposure. We use values consistent with \citet{Oza2024}: orbital period 2.78174 days, semi-major amplitude 0.0378 AU, inclination 84.48 degrees, and the RV semi-amplitude of 151.468 {\mbox{${\rm km\,s}^{-1}$}} \citep{Wyttenbach2017}. The final transmission spectra in the sodium doublet region are shown in Figure \ref{masked_residual}. To model the sodium doublet and its associated error, we assumed both lines are Gaussian, with peak absorption depth ($\Delta \mathcal{F}_{NaD}/ \mathcal{F}_{\star}$), centroid velocity ($v_{0}$) and FWHM ($\sigma$), along with a flat continuum at c as free parameters. A prior $\sigma$ value of 2.0 {\mbox{${\rm km\,s}^{-1}$}} used for the model convergence. We achieve a 8.0-$\sigma$ detection of neutral sodium corresponding to an average NaD transit depth of $\Delta \mathcal{F}_{NaD}/ \mathcal{F}_{\star} =$  3.2 $\pm$ 0.4 \% with a $D_{2}/D_{1}$ ratio of 1.4. Interestingly, the residual signal shows an overall blueshift of -6.2 $\pm$ 0.5 {\mbox{${\rm km\,s}^{-1}$}} along with a 12.4-$\sigma$ detection. The posterior distribution of the best-fit model is shown in Figure \ref{GaussianEpoch5}.

%Let's save for referee response 
%The companion paper by Oza. et.al 2024 (under review) conducted a detailed study of WASP-49b using three HARPS and one VLT/ESPRESSO transit observation. Interestingly, the observed sodium was variable. They couldn't find any sodium detection in ESPRESSO observation. They were able to observe a transit depth of $0.8 \pm 0.3 \% [2.6 \sigma]$, $3.8 \pm 0.8 \% [4.8 \sigma]$, and $0.7 \pm 0.3 \% [2.3 \sigma]$ for nights one, two, and three, respectively using HARPS observations. In addition to that, HARPS night two was showing a redshift of 9.7 $\pm$ 1.6 {\mbox{${\rm km\,s}^{-1}$}} in the sodium absorption line.

As the host star WASP-49 A is a slow rotator and less active star ($ log R_{HK} = -5.17$, \citet{Cegla_2016}), the chances of stellar contamination of the observed sodium is very unlikely. \citet{Wyttenbach2017} did a detailed study to understand the Rossiter-McLaughlin effect (RM) and found no significant impact of the RM effect on the transmission spectra at sub {\mbox{${\rm km\,s}^{-1}$}} levels. The center-to-limb variations in the sodium D lines are also not significant in the case of G-type stars like WASP-49 A \citep{Czesla_2015,Khalafinejad_2017,Wyttenbach2017}. Due to the relatively high systemic velocity (41.73 \mbox{${\rm km\,s}^{-1}$}), the stellar sodium lines are well separated from the interstellar medium lines. The ratio of the D$_2$/D$_1$ lines is generally indicative of the optical depth of the atmospheric or exospheric medium \citep{Gebek_2020}.  The KECK/HIRES data supports a more tenuous sodium exosphere $f_{D_2/D_1} \sim$ 1.4 $\pm$ 0.3, suggesting the sodium may indeed be at the border of an optically thin/optically thick medium ($\tau_{\nu} \sim 0.7 - 3 $). Figure \ref{DopplerCC} shows the residual sodium spectra for the D$_2$ line as a function of radial velocity at different orbital phases during transit. Flat spectra before and after transit (black) confirm the insignificant stellar contamination. A blue-to-red shift can be seen as we move from ingress (blue) to egress (green) of the transient feature. We combine three exposures centered at $\phi_{exoplanet} = -0.0076$ referred to as ingress, three exposures centered at $\phi_{exoplanet} = -0.0018$ exoplanet mid-transit, and three exposures centered at $\phi_{exoplanet} = +0.0076$ referred to as egress of the transient sodium feature. The observed Na I Doppler shifts are distributed blueward, with an average shift of -6.2 $\pm$ 0.5 {\mbox{${\rm km\,s}^{-1}$}} of the planetary rest frame, even though the residual at egress is surprisingly redshifted at $\sim$ +4.1 $\pm$ 1.5 {\mbox{${\rm km\,s}^{-1}$}} and the ingress is blueshifted at roughly -10.3 $\pm$ 1.9 {\mbox{${\rm km\,s}^{-1}$}}. We confirm variable Doppler behavior at high-significance with $>$5$\sigma$ blueshifts, and $\sim$ 3$\sigma$ redshifts, reported at +9.7 $\pm$ 1.6 {\mbox{${\rm km\,s}^{-1}$}} in HARPS/3.6-m Epoch II data \citep{Oza2024}. 
% recently reported an average redshift  and a marginal blueshift in two HARPS observations of WASP-49b, along with a non-detection of sodium in one HARPS and one ESPRESSO observation.

\begin{figure}%[htp]
\centering
\includegraphics[width=1.0\columnwidth]{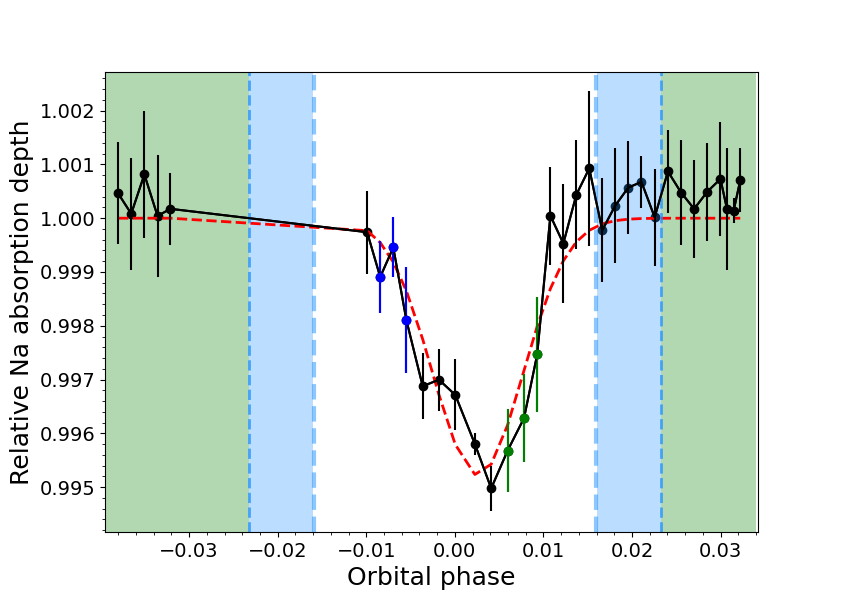}
\caption{Average neutral sodium lightcurve constructed using 1.5 {\AA} band width from the core of both Na D$_{1}$ and Na D$_{2}$ lines. The white region indicates the transit duration, the blue region is $\sim \frac{1}{2}$ the Hill sphere, and the green region indicates the out-of-transit observation. The estimated transit duration of the sodium signal from the best-fit Gaussian (red dashed line) is 40.05 $\pm$ 3.2 minutes. The average value of the blue and green data points used as ingress and egress in Figure \ref{DopplerCC}, respectively. } \label{HIRES_lc}
\end{figure}

To confirm the origin, we further analyzed the sodium signature as a function of time \citep{Seidel2019,Oza2024}. We constructed a narrow band lightcurve by averaging sodium D lines of 1.5 {\AA} width centered around each line. Figure \ref{HIRES_lc} shows the average sodium transit lightcurve of WASP-49b at $d\mathcal{F}/\mathcal{F} (t) \sim 0.47\% \pm 0.04 \% (11.7 \sigma$). 
Compared to the 129-minute exoplanet transit window, we find a Gaussian FWHM duration of $\Delta t_{NaD} \sim$ 40.05 minutes (average residual duration over $>$3-sigma), roughly 90 minutes shorter than the exoplanet transit duration. Interestingly, the duration of the transient sodium in the lightcurve is also reported to be roughly $\sim$ 40 minutes by \citet{Oza2024}, which we confirm here at 11.7 $\sigma$.
%{\bf Unfortunately, a couple of exposures are missing ($-0.03 \leq \phi_{exoplanet} \leq -0.012 $) close to the ingress ($\phi_{exoplanet} = -0.0016$) which makes this difficult to interpret.} 
The velocity asymmetry shown in Figure \ref{DopplerCC} is different from kinematic behavior usually observed in other hot Jupiters \citep{Ehrenreich_2020,Kesseli_2022}. For a hot Jupiter, generally a redshift or no velocity shift is measured at ingress followed by a blueshift at egress, which is expected if the signal is coming from a planet rotating synchronously in the direction of its orbit. A signal that is more blueshifted at ingress could indicate the observed sodium is not from the atmospheric limb of the planet. The magnitude of the velocity shift is also similar to an ultra-hot (T$_{eq}$ $\gtrsim$ 2000 K) planetary jet escaping from an exoplanet atmosphere \citep{Seidel_2023}, where the kinetic temperatures probed at 1 nanobar has been reported to exceed $\sim$ 12,000 K \citep{huang23}.  
%{\bf Here, the equilibrium temperature of WASP-49b is 1400 K \citep{wyttenbach17}. Furthermore, although day-night temperature gradients, and radiation pressure ($\sim$ - 5.7 {\mbox{${\rm km\,s}^{-1}$}}) can drive this observed blueshift.}
%, a $\gtrsim +$ 10 {\mbox{${\rm km\,s}^{-1}$}} redshift at egress is surprising.}
Multi-epoch sodium lightcurve observations thus far of this hot Saturn indicate the observed sodium is Doppler-shifted and transient, which we confirm here, likely associated with the orbit of a natural satellite  \citep{Oza2024}. Recent dynamical studies demonstrate low-eccentricity exomoons are indeed stable at orbits in between 1.2-1.5 R$_p$ \citep{sucerquiacuello2025}.  Dozens of more large-aperture, precise high-resolution transmission spectroscopy observations are needed to build up a radial velocity follow up of the Doppler shift presented here.  High- cadence infrared observations of volcanic gases (SO$_2$, CO$_2$) or grains (SiO$_2$) would be especially useful in tandem with high-resolution optical observations.
%All four observations also revealed a variable sodium light curve. The keck/HIRES results support the transient nature of the sodium observed most probably from a co-orbiting natural satellite.
%At high altitudes, endogenic Na I is unlikely to survive throughout the entirety of the $\sim$ 129 minute transit duration, at an orbit of only $\sim$7.8 R$_{\star}$ to the host Sun-like star \citep{Huebner_2015}. The photoionization timescale is very short $\sim$ 4 minutes, only $\sim$ 3\% of the transit duration, which requires a sodium supply as suggested by \citet{oza_2019}. 
%\textbf{ Thus far in sodium light, the Doppler blueshift to redshift detected here from -14.4 km/s to +7.4 km/s in Fig.6 is highly suggestive of a natural satellite orbit \citet{Oza2024}, with allowed orbital velocities ranging from -15 to +15 km/s bounded between the Roche and Hill spheres \citet{SucerquiaCuello2025}. If the $\sim$40 minute feature confirmed here continues to repeat, transit geometry (Lecavelier des Etangs et al. 2007) concludes a semi-major axis of 1.4 R$_p$. Of course, if the NaD is moreso indicative of the plasma processes of the natural satellite and planetary magnetosphere \citep{MeyerzuWestram2024} then the Doppler RV measurements would rather indicate the B field of the exoplanet \citep{narang2023a, narang2023a}, both of which require dozens of high-resolution transit observations to confirm.} 
%%%%%%%%%%%%%%%%%%%
\section{Conclusion} \label{section:4}
%%%%%%%%%%
In summary, we observed one complete transit of the hot Saturn WASP-49 b, including significant stellar baseline using Keck/HIRES. Using a custom wavelength calibration technique, we were able to achieve a wavelength precision of 60 {\mbox{${\rm m\,s}^{-1}$}}. This work demonstrates the feasibility of using slit-based spectrographs Keck/HIRES for high-resolution exoplanet transit spectroscopy.  Here, we identify an average residual sodium flux of $d\mathcal{F}_{Na}/\mathcal{F}_{\star} (\lambda)=$ 3.2 $\pm$ 0.4 \%  in transmission spectra (8.0 $\sigma$), with an observed Doppler shift of -6.2 $\pm$ 0.5 {\mbox{${\rm km\,s}^{-1}$}} suggests the neutral sodium gas is moving with respect to the exoplanet. The approximate flux of the transient feature in the transmission lightcurve is: $d\mathcal{F}_{Na}/\mathcal{F}_{\star} (t)=$ 0.47 $\pm$ 0.04 \% which is smaller than the HARPS/3.6-m \citep{Oza2024}. The kinematic Na I results indicate that the source of the observed sodium is unlikely from the atmosphere of the planet, yet from a natural satellite orbiting around the planet, although we cannot be sure of its orbit or the geometry of its clouds based on this transit alone. 
\section*{Acknowledgements}
We acknowledge the Keck telescope facility and archive facility for providing the observation time and reduced data. We would like to acknowledge the Exo-Io team for all the support and help. The research described in this paper was carried out in part at the Jet Propulsion Laboratory, California Institute of Technology, under a contract with the National Aeronautics Space Administration. © 2025. All rights reserved. Some of the data presented herein were obtained at Keck Observatory, which is operated as a scientific partnership among the California Institute of Technology, the University of California, and the National Aeronautics and Space Administration. The Observatory was made possible by the generous financial support of the W. M. Keck Foundation. The authors wish to recognize and acknowledge the very significant cultural role and reverence that the summit of Maunakea has always had within the Native Hawaiian community. We are most fortunate to have the opportunity to conduct observations from this mountain.

%a private 501(c)3 non-profit organization 
%%%%%%%%%%%%%%%%%%%%%%%%%%%%%%%%%%%%%%%%%%%%%%%%%%
\section*{Data Availability}
The observed KECK/HIRES data is available on the Keck Observatory Archive (\href{https://www2.keck.hawaii.edu/koa/public/koa.php}{KOA}).

\bibliographystyle{mnras}
\bibliography{ref} 

\appendix
\renewcommand{\thefigure}{A\arabic{figure}} % This will prefix 'A' to the figure number
\setcounter{figure}{0} % Reset the figure counter to start from 1 in the appendix
\begin{figure}
\begin{center}
\includegraphics[width=1.0\columnwidth]{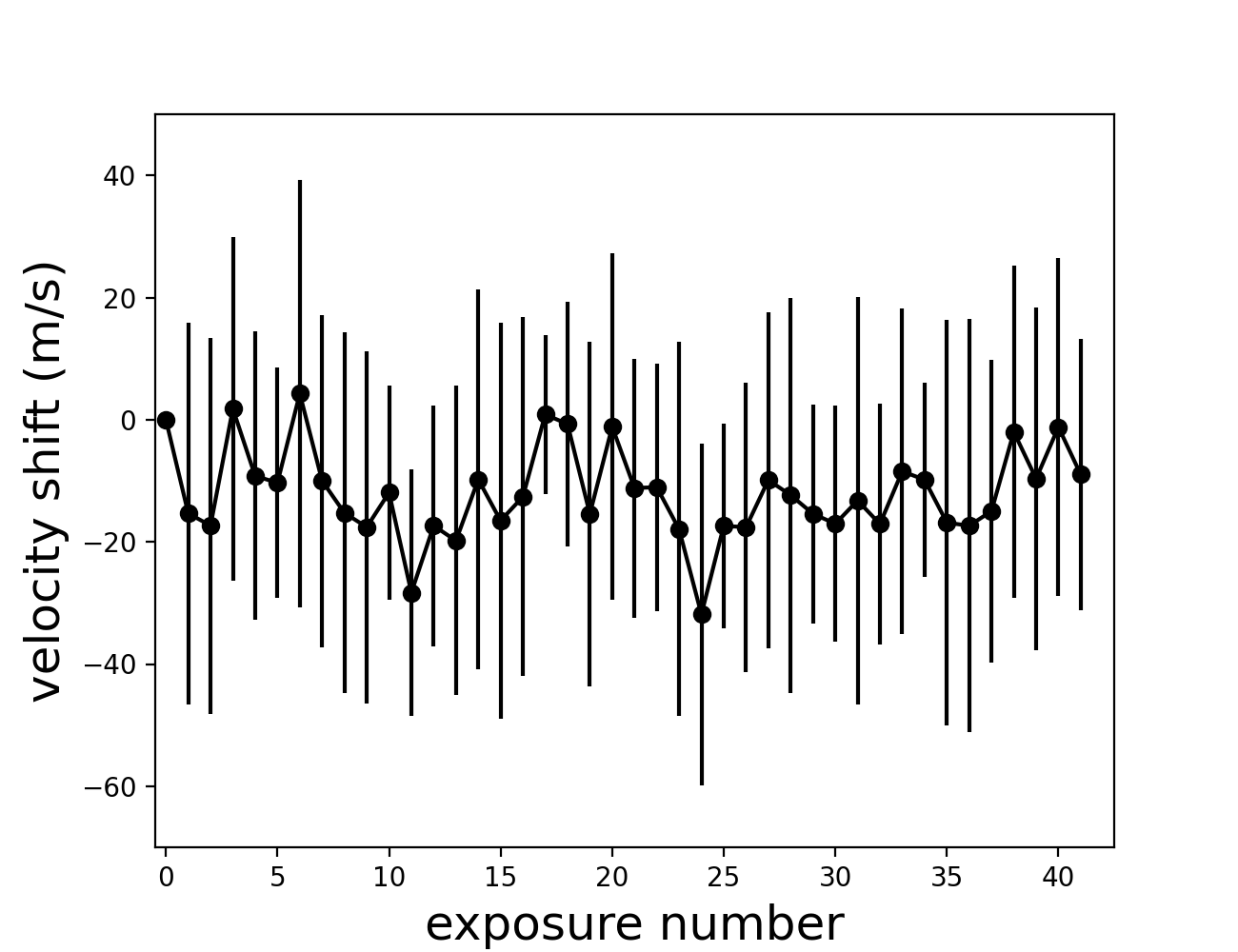}
\caption{The velocity precision of the wavelength solution acheived by Keck/HIRES after distortion correction for the echelle order with sodium. Each data point is the average wavelength shift in all the clean absorption line in an exposure.} \label{distortion}
\end{center}
\end{figure}

\begin{figure}
\begin{center}
\includegraphics[width=1.0\columnwidth]{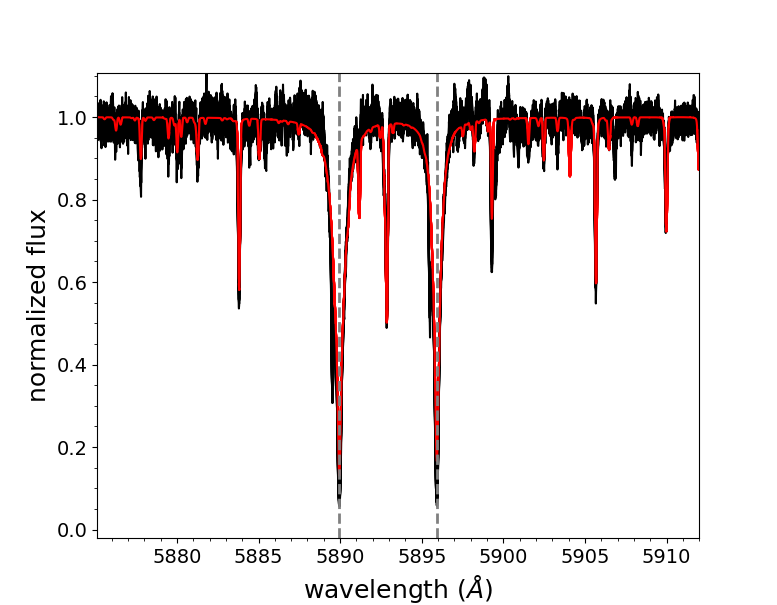}
\caption{All the HIRES exposures (black) overplotted with the Phoenix model in the Stellar Rest Frame. The vertical dashed lines (grey) show the position of Na D lines in the Stellar Rest Frame.} \label{spectra_overplot}
\end{center}
\end{figure}

\begin{figure}
\begin{center}
\includegraphics[width=1.0\columnwidth]{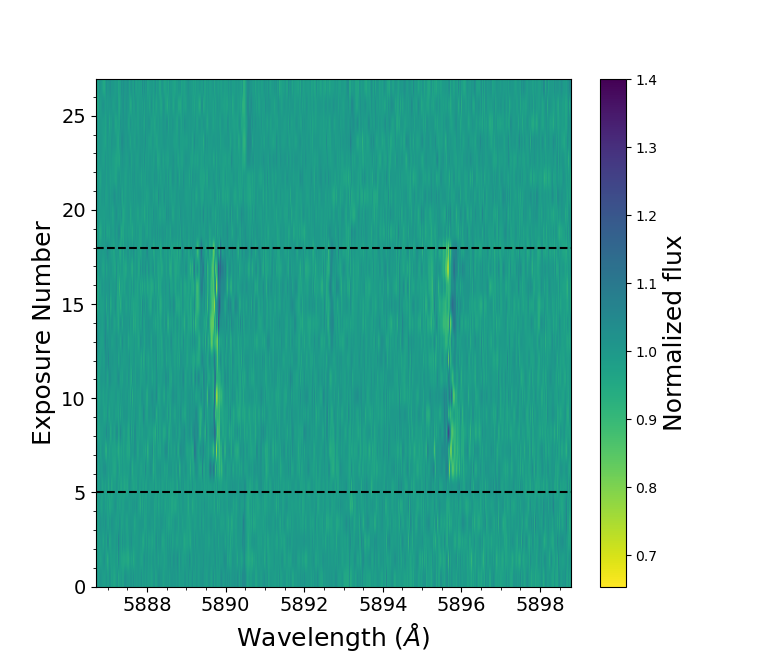}
\caption{Doppler-shifted sodium detection in the Stellar Rest Frame (SRF). The region within the horizontal lines shows the exoplanet transit event. The colorbar indicates the normalized flux. } \label{matrix_plot}
\end{center}
\end{figure}

\bsp	% typesetting comment
\label{lastpage}
\end{document}